\documentclass[letterpaper]{jpconf}
\usepackage{graphicx}
\begin{document}
\title{Present status of the nonstrange and other flavor
       partners of the exotic $\Theta^+$ baryon}

\author{I.~I.~Strakovsky$^1$,
R.~A.~Arndt$^1$,
Ya.~I.~Azimov$^2$,
M.~V.~Polyakov$^{2,}$$^{3,}$$^4$,
and R.~L.~Workman$^1$}

\address{$^1$Center for Nuclear Studies, Department of Physics,
    The George Washington University, Washington, D.C.
    20052, USA}
\address{$^2$Petersburg Nuclear Physics Institute, Gatchina,
    St.~Petersburg 188300, Russia}
\address{$^3$Universit\'e  de Li\`ege au Sart Tilman,
    B-4000 Li\`ege 1 Belgium}
\address{$^4$Institute for Theoretical Physics~II, Ruhr University,
    44780 Bochum, Germany}

\ead{igor@gwu.edu}

\begin{abstract}
Given the existing empirical information about the exotic
$\Theta^+$ baryon, we analyze possible properties of its
$SU(3)_F$-partners, paying special attention to the
nonstrange member of the antidecuplet $N^\ast$. The
modified $\pi N$ partial-wave analysis presents two
candidate masses, 1680~MeV and 1730~MeV.  In both cases,
the $N^\ast$ should be rather narrow and highly inelastic.
Our results suggest several directions for experimental
studies that may clarify properties of the antidecuplet
baryons, and structure of their mixing with other baryons.
Recent experimental evidence from the GRAAL and STAR
Collaborations could be interpreted as observations of
a candidate for the $\Theta^+$ nonstrange partner.
\end{abstract}.

Results from a wide range of recent experiments are 
consistent with the existence of an exotic S=+1 resonance, 
the $\Theta^+(1540)$ with a narrow width and a mass near 
1540~MeV~\cite{Nakano}.  Direct width determinations have been 
hindered by the limitations of experimental resolution, 
resulting in upper bounds of order 10~MeV.  The quantum 
numbers of this state remain unknown, though a prediction 
of $J^P$ = $1/2^+$ was obtained in the work~\cite{DPP} 
that provided motivation for the original search.

\begin{figure}[ht]
\includegraphics[width=16pc, angle=90]{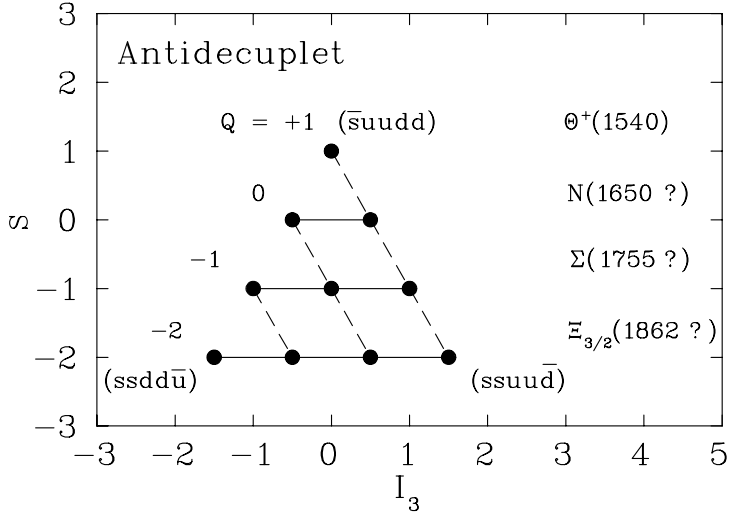}\hspace{2pc}%
\begin{minipage}[b]{13pc}\caption{\label{fig:g1}Tentative unitary 
       anti-decuplet with $\Theta^+$.  Isotopic multiplet (constant 
       values of the charge) shown by solid (dashed) lines.}
\end{minipage}
\end{figure}

Additional information related to the assignment of unitary 
partners is due to a recent experimental result~\cite{Xi} 
giving evidence for one further explicitly exotic particle 
$\Xi^{--}_{3/2}$, with a mass $1862\pm2$~MeV and width $< 
18$~MeV ({\it i.e.} less than resolution).  Such a particle 
had been expected to exist as a member of an
antidecuplet, together with the $\Theta^+$.  However, the soliton 
calculation of the mass difference requires some assumptions.  
In particular, it depends on the value of the $\sigma$-term, 
which is the subject of controversy.  Its value, taken
according to the latest $\pi N$ data analysis~\cite{ABSWP}, 
leads to an antidecuplet mass difference of about 110~MeV,
instead of the originally predicted 180~MeV~
\cite{DP}.  So, if the states $\Xi_{3/2}$~\cite{Xi} and 
${\Theta}$ are indeed members of the same antidecuplet, then, 
according to the Gell-Mann-Okubo rule, the mass difference of 
any two neighboring isospin multiplets in the antidecuplet 
should be constant and equal $(M_{\Xi_{3/2}} - 
M_{\Theta})/3\approx 107~{\rm MeV,}$ which fits the GW SAID 
$\sigma$-term result very well~\cite{ABSWP}.  This shift also effects
the masses of other unitary partners of 
the $\Theta^+$: nucleon-like and $\Sigma$-like.  The 
supposed antidecuplet,with $\Sigma$- and $N$-masses determined by 
the Gell-Mann--Okubo rule, looks today as shown on Fig.~\ref{fig:g1}.

Due to $SU(3)_F$-violating mixing with lower-lying nucleon-like 
octet states, $M_{N^\ast}$ may shift upward, and reach 
about 1680~MeV~\cite{DP}.  Mixing with higher-lying 
nucleon-like members of exotic 27- and 35-plets may also 
play a role.  

The state $N(1710)$, though listed in the PDG Baryon
Summary Table~\cite{PDG04} as a 3-star resonance and used
as input in the $\Theta^+$ prediction~\cite{DPP}, is not 
seen in the latest analysis of pion-nucleon elastic scattering 
data (see Table~\ref{tbl:tbl1}).  Studies which 
have claimed to see this state have given widely varying 
estimates of its mass and width (from $\sim 1680$~MeV to 
$\sim 1740$~MeV for the mass and from $\sim 90$~MeV to 
$\sim 500$~MeV for the width).  Branching ratios have also 
been given with large uncertainties (10--20\% for $N\pi$, 
40--90\% for $N\pi\pi$, and so on), apart from one which has 
been presented with much greater precision ($6\pm 1\%$ for 
$N\eta$).  In any case, the PDG width of $N(1710)$ seems to 
be too large for the partner of the narrow $\Theta^+$.
It would be more natural for members of the same unitary 
multiplet to have comparable widths.

As has been suggested recently (see Refs.~\cite{AASW,AAPSW}), 
any standard PWA by itself tends to miss narrow resonances.  
For this reason, we considered~\cite{AASW} 
a modified PWA,
assuming the existence of a narrow resonance, and compared the
quality of fit with and without such structures (more 
detailed description of formulas see in Ref.~\cite{AAPSW}). 
Such an approach was used initially to look for light 
nucleon resonances~\cite{AASW}.

This method, applied to studies of the $\Theta^+(1540)$~
\cite{ASW}, though unable to determine the $\Theta^+$ 
quantum numbers, placed a tight limit on the width, in full agreement 
with the results of other approaches.  We subsequently used this 
method~\cite{AAPSW} to search $\pi N$ scattering data for a 
narrow nucleon-like state assumed to be a member of the 
antidecuplet, accompanying the $\Theta^+(1540)$.  The two 
candidate masses, M$_R$ = 1680~MeV and 1730~MeV, would necessary 
be quite inelastic
with $\Gamma_{el} <$ 0.5~MeV and 0.3~MeV, respectively.
Some support for a narrow structure in this mass region has
recently been obtained in a preliminary report based on
direct measurements by the STAR~\cite{KAB} and GRAAL~\cite{SLAV}
Collaborations.  

It should be noted that there have been other publications finding no
evidence for the existance of the $\Theta^+$.  These findings need more
detailed consideration.  Though the present non-observation data 
require some suppression of exotic production, as compared to
conventional hadrons, they do not definitively exclude the existence of the
$\Theta^+$ and/or its companions/analogs.
For example, analysis of the BES data~\cite{BES} in Ref.~\cite{AS} 
shows that indeed there is some suppression of $\Theta$-production.
However, suppression of the exotics at this level could have
a quite natural explanation.

There is also a statement that the observed peak of $\Theta^+$
could be due to a kinematical reflection of some of known
resonances.  A particular consideration 
has been suggested by 
Dzierba {\it et al.}~\cite{Dzierba} addressed to the CLAS 
analysis~\cite{CLAS}.  The specific model used by Dzierba
{\it et al.} has, however, 
been criticized~\cite{DzierbaS}, and may not be a serious
concern for the CLAS results.

We should emphasize here that if the present evidence 
for the $\Theta$ turns out to be incorrect, we would 
have to answer a different, but also difficult, question:
why do we {\it not} see exotic hadrons?
We take here, for the sake of argument, the position that the $\Theta$ does 
exist, but its production is governed by different mechanisms.
Though we essentially agree with suggestions of 
Karliner and Lipkin~\cite{KarLip} about how to clarify the 
problem, we think that, first of all, it is especially 
important to reliably confirm the existence of the $\Theta$ 
in the processes where it has been reported. 
New data are being collected for this purpose,
by several collaborations, and one would hope for a definitive
answer within a year.

That is why, at the moment, we will assume the $\Theta^+$ 
(as well as other multi-quark hadrons) being existent, and 
will discuss some consequences of this fact (for details,
see Ref.~\cite{AAPSW}).

To summarize, given our current knowledge of the $\Theta^+$, 
the state commonly known as the $N(1710)$ is not the 
appropriate candidate to be a member of the antidecuplet 
together with the $\Theta^+$.  Instead, we suggest candidates 
with nearby masses, $N(1680)$ (more promising) and/or $N(1730)$ 
(less promising, but not excluded).  Our analysis suggests 
that the appropriate state should be rather narrow and very
inelastic.  Similar considerations have been applied to the 
$\Xi_{3/2}(1862)$, assumed to be also a member of the same 
antidecuplet.  It should also be quite narrow.

How reliable are our theoretical predictions?
They have, indeed, essential theoretical uncertainties.  
We have yet to establish the existence of the (narrow)
state originally associated with the $N(1710)$.
Moreover, we have assumed the presence of only one state with $J^P 
= 1/2^+$, either $N(1680)$ or $N(1730)$.  If both exist
with the same spin and parity, our conclusions should be 
reconsidered.

Futhermore, we use the mixing angle $\phi$, taken from 
Ref.~\cite{DPP}, which was actually determined through formulas 
containing the $\sigma$-term (just as the mass difference in the 
antidecuplet). 
If we use parameters corresponding to more recent information,
for both the $\sigma$-term and the mass difference, we obtain 
larger mixing, up to $\sin\phi\approx 0.15$.  With our formulas, 
this would most strongly influence the partial width 
$N^{\ast}\to\pi\Delta$, increasing it to about 15~MeV.  Other 
partial widths of $N^{\ast}$ change not so dramatically, and 
the total width appears to remain not higher than $\sim 30$~MeV.  
Such a width could well be measured, but not in elastic
scattering, because of an expected very small elastic branching 
ratio.  Note, however, that the above large value for $\sin\phi$ 
may appear problematic, since the formulas of Ref.~\cite{DPP} 
assume linearisation with respect to $SU(3)_F$-violation, and 
need to be reconsidered if the violation appears to be large.

Nevertheless, even having in mind all theoretical uncertainties, 
we can suggest several directions for experimental studies.  
First of all, one should search for possible new narrow nucleon 
state(s) in the mass region near 1700~MeV.  Searches 
may use various initial states, ({\it e.g.}, $\pi N$ collision 
or photoproduction).  We expect the largest effect in the 
$\pi\pi N$ final state (mainly through $\pi\Delta$, though it is 
forbidden by $SU(3)_F$).  The final states $\eta N$ and 
$K\Lambda$ may also 
be interesting and useful, especially the ratio of $\eta N$ 
and $\pi N$ partial widths, as the latter is very sensitive to 
the structure of the octet--antidecuplet mixing.  Another 
interesting possibility to separate antidecuplet and octet 
components of $N^\ast$ is provided by comparison of photoexcitation 
amplitudes for neutral and charged isocomponents of this
resonance, the point being that the antidecuplet component 
contribution to the photoexitation of the charged $N^\ast$
is strongly suppressed (see details in Ref.~\cite{Polyakov}).  

On the other hand, such a relatively simple picture of mixing can not 
reproduce our small value(s) of $\Gamma_{el}$.  We assumed in our
analysis that this could result from more complicated mixing with 
several other multiplets~\cite{AAPSW}.  Such a possibility was 
recently confurmed~\cite{PG}.

For $\Xi_{3/2}$, attempts to measure the total width are necessary,  
though it could possibly be even smaller than $\Gamma_{\Theta^+}$.  
Branching ratios for $\overline K\Sigma$ and $\pi\Xi(1530)$, in 
relation to $\pi\Xi$, are very interesting.  These may give
important information on the mixing of antidecuplet baryons with 
octets and higher $SU(3)_F$-multiplets.

\subsection{Acknowledgments}
The work was partly supported by the U.~S.~Department of 
Energy Grant DE--FG02--99ER41110, by the Jefferson Laboratory, 
by the Southeastern Universities Research Association under 
DOE Contract DE--AC05--84ER40150, by the Russian State Grant 
SS--1124.2003.2.


\smallskip
\begin{table}[ht]
\caption{\label{tbl:tbl1}
         Comparison of N(1710) properties.}
\begin{tabular}{cccc}
\br
Collaboration  & Mass (MeV)      &Width (MeV)& Ref \\
\mr
DPP            & 1710 (input)    &$<$40      &\protect\cite{DPP}  \\
\mr
KH             & 1723$\pm$9      &120$\pm$15 &\protect\cite{KH}   \\
CMU            & 1700$\pm$50     & 90$\pm$30 &\protect\cite{CUT}  \\
KSU            & 1717$\pm$28     &480$\pm$230&\protect\cite{MAN}  \\
GWU            & 1710            & no state !&\protect\cite{ABSWP}\\
\br
\end{tabular}
\end{table}

\section{References}

\medskip
\begin{thebibliography}{99}
\bibitem{Nakano} See for example, Nakano T 2004, Plenary talk at 
      NSTAR 2004, Grenoble, March 2004
\bibitem{DPP}Diakonov D, Petrov V and Polyakov M 1997 {\it Z. 
      Phys. A}{\bf 359} 305
\bibitem{Xi}Alt C {\it et al.} 2004 {\it Phys. Rev. Lett.} {\bf 92} 
      042003
\bibitem{ABSWP}Arndt R A, Briscoe W J, Strakovsky I I, Workman R L,
      and Pavan M M 2004 {\it Phys. Rev. C} {\bf 69} 035213
\bibitem{DP}Diakonov D and Petrov V, 2004 {\it Phys. Rev. D}
      {\bf 69} 094011
\bibitem{PDG04}Eidelman S \textit{et al.} 2004 {\it Phys. Lett. B}
      {\bf 592} 1
\bibitem{KH}Koch R 1985 {\it Z. Phys. C} {\bf 29}, 597;
      H\"ohler G 1983 {\it Pion--Nucleon Scattering},
      Landoldt--B\"ornstein Vol. \textbf{I/9b2}, edited by Schopper
      H (Springer Verlag).
\bibitem{CUT}Cutkosky R E {\it et al.} 1980 in {\it Proceedings of
      4th International Conference on Baryon Resonances, Toronto,
      Canada, July 1980}, published in Baryon 1980:19
      (QCD161:C45:1980); Cutkosky E E and Wang S 1990 {\it Phys.
      Rev. D} {\bf 42} 235
\bibitem{MAN}Manley D M and Saleski E M 1992 {\it Phys. Rev. D}
      {\bf 45} 4002
\bibitem{AASW}Azimov Ya I, Arndt R A, Strakovsky I I, and Workman
      R L 2003 {\it Phys. Rev. C} {\bf 68} 045204
\bibitem{AAPSW}Arndt R A, Azimov Ya I, Polyakov M V, Strakovsky I I,
      and Workman R L 2004 {\it Phys. Rev. C} {\bf 69} 035208
\bibitem{ASW}Arndt R A, Strakovsky I I, and Workman R L 2003 {\it
      Phys. Rev. C} {\bf 68} 042201
\bibitem{KAB}Kabana S, hep-ex/0406032
\bibitem{SLAV}Kuznetsov V, hep-ex/0409032
\bibitem{BES}Bai J Z {\it et al.} 2004 {\it Phys. Rev. D} {\bf 70}
      012004
\bibitem{AS}Azimov Ya I and Strakovsky I I 2004 {\it Phys. Rev. C}
      {\bf 70} 035210
\bibitem{Dzierba}Dzierba A {\it et al.} 2004 {\it Phys. Rev. D}
      {\bf 69} 051901
\bibitem{CLAS}Stepanyan S {\it et al.} 2003 {\it Phys. Rev. Lett.}
      {\bf 91} 252001
\bibitem{DzierbaS}Hicks K, Burkert V, Kudryavtsev A, Strakovsky I,
      and Stepanyan S, Submitted to {\it Phys. Rev. D}
      hep-ph/0411265
\bibitem{KarLip}Karliner M and Lipkin H 2004 {\it Phys. Lett. B}
      {\bf 597} 309
\bibitem{Polyakov}Polyakov M V and Rathke A 2003 {\it Eur. Phys.
       J. A} {\bf 18} 691
\bibitem{PG}Gusey V and Polyakov M V, hep-ph/0501010
\end{thebibliography}
\end{document}